\documentclass{PoS}

\usepackage{graphicx}
\usepackage{subfigure}
\usepackage{amsmath}

\title{Photon-induced contributions to di-lepton production at the LHC Run II}

\ShortTitle{Photon-induced contributions to di-lepton production}

\author{\speaker{Juri Fiaschi}\\
        School of Physics \& Astronomy, University of Southampton\\
        E-mail: \email{juri.fiaschi@soton.ac.uk}}

\author{Elena Accomando\\
       School of Physics \& Astronomy, University of Southampton\\
       E-mail: \email{e.accomando@soton.ac.uk}}
       
\author{Francesco Hautmann\\
       Rutherford Appleton Laboratory and University of Oxford\\
       University of Antwerp, Elementary Particle Physics\\
       E-mail: \email{hautmann@thphys.ox.ac.uk}}
       
\author{Stefano Moretti\\
       School of Physics \& Astronomy, University of Southampton\\
       E-mail: \email{s.moretti@soton.ac.uk}}

\author{Claire H. Shepherd-Themistocleous\\
       Particle Physics Department, Rutherford Appleton Laboratory\\
       E-mail: \email{claire.shepherd@stfc.ac.uk}}

\abstract{We report on recent studies of photon-induced (PI) contributions 
to di-lepton production and their implications for Beyond Standard Model (BSM) $Z^\prime$-bosons searches at the LHC.}

\FullConference{XXV International Workshop on Deep-Inelastic Scattering and Related Subjects\\
		3-7 April 2017\\
		University of Birmingham, UK}

\begin{document}

\section{Introduction}

Di-lepton final states in the high invariant mass region 
$ M_{l l } \geq 1 $ TeV are one of the primary channels used in 
searches for $Z^\prime$ gauge bosons in Beyond Standard Model (BSM) scenarios and in 
 precision studies of the Standard Model (SM) at the Large Hadron Collider 
(LHC). 

It was pointed out in~\cite{Accomando:2016tah, Accomando:2016ouw} that in the high-mass region the contribution of photon-induced (PI) 
di-lepton production, and associated uncertainties from the photon parton distribution function (PDF), 
could affect the di-lepton SM spectrum shape potentially influencing wide-resonance searches. 
This could have an even greater impact on the Contact Interaction (CI) type of search, where one has to predict the SM background from theory. 
Therefore, while these effects do not influence significantly standard narrow $Z^\prime$ searches, they can become a significant source 
of theoretical systematics for wide $Z^\prime$-bosons and CI searches.

The analysis of this theoretical systematics has been extended in~\cite{Accomando:2016ehi,Accomando:2017itq} by, on one hand,
including the contribution of both real and virtual photon processes and, on the other hand, evaluating the impact of recent updates in photon PDF fits. 

This article gives a brief report on these studies. 

\section{Real and virtual photon processes in di-lepton production}

The contribution of photon interactions from elastic collisions is generally evaluated through the Equivalent Photon Approximation (EPA), following 
established literature~\cite{Budnev:1974de}.
With the introduction of QED PDFs, also inelastic processes can be included in the picture and their contribution can be theoretically 
estimated.
In the QED PDF framework indeed, the initial resolved photons (i.e. real photons with $Q^2=0$) have their own PDF which describes their
initial state.

Examples of inelastic QED PDF sets are the CT14QED~\cite{Schmidt:2015zda} and the MRST2004QED~\cite{Martin:2004dh} sets.
These inelastic sets can be used to calculate separately the three contributions coming from two resolved photons, one resolved and one virtual, 
and two virtual photon scattering.
These three terms are called Double-Dissociative (DD), Single-Dissociative (SD) and pure EPA respectively.
The implementation of the virtual photon spectrum and the choices of its parameters follow the work in Ref.~\cite{Piotrzkowski:2000rx}.

\begin{figure}
\centering
\includegraphics[width=0.55\textwidth]{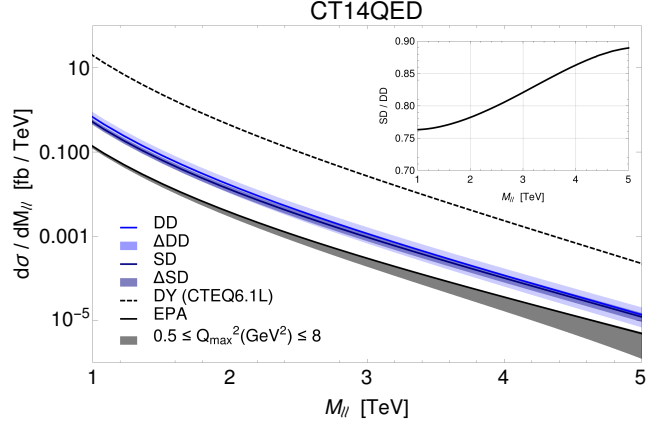}
\caption{\footnotesize Predictions for the DD, SD and pure EPA terms obtained with the CT14QED inelastic
QED PDF sets, in comparison with the dominant DY contribution.}
\label{fig:DD_SD}
\end{figure}

The results for the contributions of 
these three terms to the dilepton cross section, evaluated using the CT14QED set, are visible in Fig.~\ref{fig:DD_SD}.
The DD contribution is generally dominant, but the SD can be equally important.
In the inset plot is shown its relative size compared with the DD result.
In the CT14QED framework the SD term contribution is about 75\% - 90\% of the DD one.
The pure EPA contribution instead is sub-dominant with respect to the other terms.

PDF collaborations also release inclusive QED PDF sets, where both the elastic and inelastic components are contained in the photon PDF
fitting procedure.
To this group belong the NNPDF3.0QED~\cite{Ball:2014uwa}, xFitter\_epHMDY~\cite{Giuli:2017oii}, LUXqed~\cite{Manohar:2016nzj} and
CT14QED\_inc~\cite{Schmidt:2015zda} PDF sets.

\begin{figure}
\centering
\includegraphics[width=0.55\textwidth]{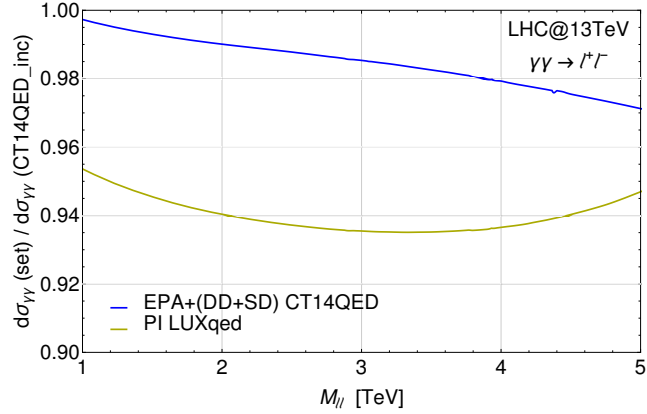}
\caption{\footnotesize Ratio between the sum of the DD, SD and pure EPA terms 
 obtained with the CT14QED set and the inclusive result 
obtained with the CT14QED\_inc set (blue line), and between the inclusive LUXqed result and the CT14QED\_inc set (yellow line).}
\label{fig:CT14QED_LUXqed}
\end{figure}

The sum of the three contributions in 
Fig.~\ref{fig:DD_SD} 
can be compared with the full results that are obtained invoking inclusive QED PDFs.
Using the CT14QED set and its inclusive version, we can directly compare the PI di-lepton spectra obtained in the two frameworks.
This comparison is visible in the blue line of Fig.~\ref{fig:CT14QED_LUXqed} which represent the ratio between the sum of the EPA, SD and DD terms
evaluated with the CT14QED, and the inclusive CT14QED\_inc result.
The difference between the two results is below 3\% in the explored invariant mass region.
This ensures that the separation between elastic and inelastic components is well understood and double counting is kept at the percent level.

In the same figure, we considered also the inclusive result that is obtained with the LUXqed set.
The yellow line shows the ratio between the two inclusive PI predictions obtained respectively with the LUXqed and with the CT14QED\_inc sets.
The two predictions are in good agreement, their difference resulting below 7\% in the explored invariant mass region.

\section{PDF uncertainties}

In Fig.~\ref{fig:DD_SD} we have given a first indication of the size of the PDF errors, which are represented by the shaded areas in the DD and SD curves.
In this section we discuss more in detail the systematic uncertainties on the PI predictions due to the photon PDF.

Generally PDF packages are accompanied by tables of PDFs, which can be used to estimate the uncertainties on their fits.
The most common approaches follow either the ``replicas'' method (NNPDF3.0QED, xFitter\_epHMDY)~\cite{Ball:2011gg} or the Hessian method 
(LUXqed)~\cite{Butterworth:2015oua}.
The CT14QED\_inc set and its inelastic version instead include a table of PDFs that are obtained imposing an increasing constraint on the fraction of 
proton momentum carried by the photon and the PDF uncertainty can be estimated following the indications in Ref.~\cite{Schmidt:2015zda}.

\begin{figure}
\centering
\includegraphics[width=0.47\textwidth]{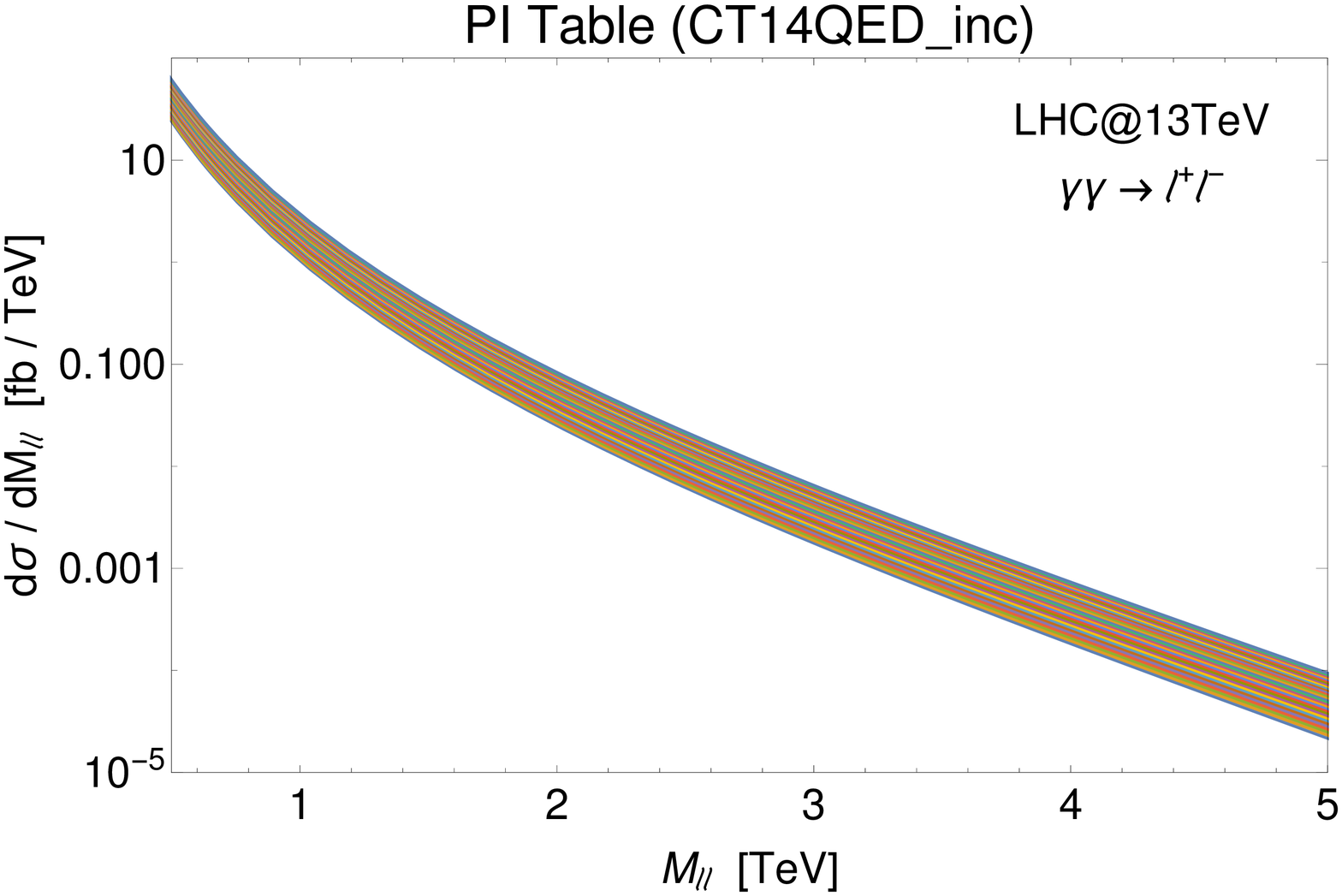}{\footnotesize (a)}
\includegraphics[width=0.47\textwidth]{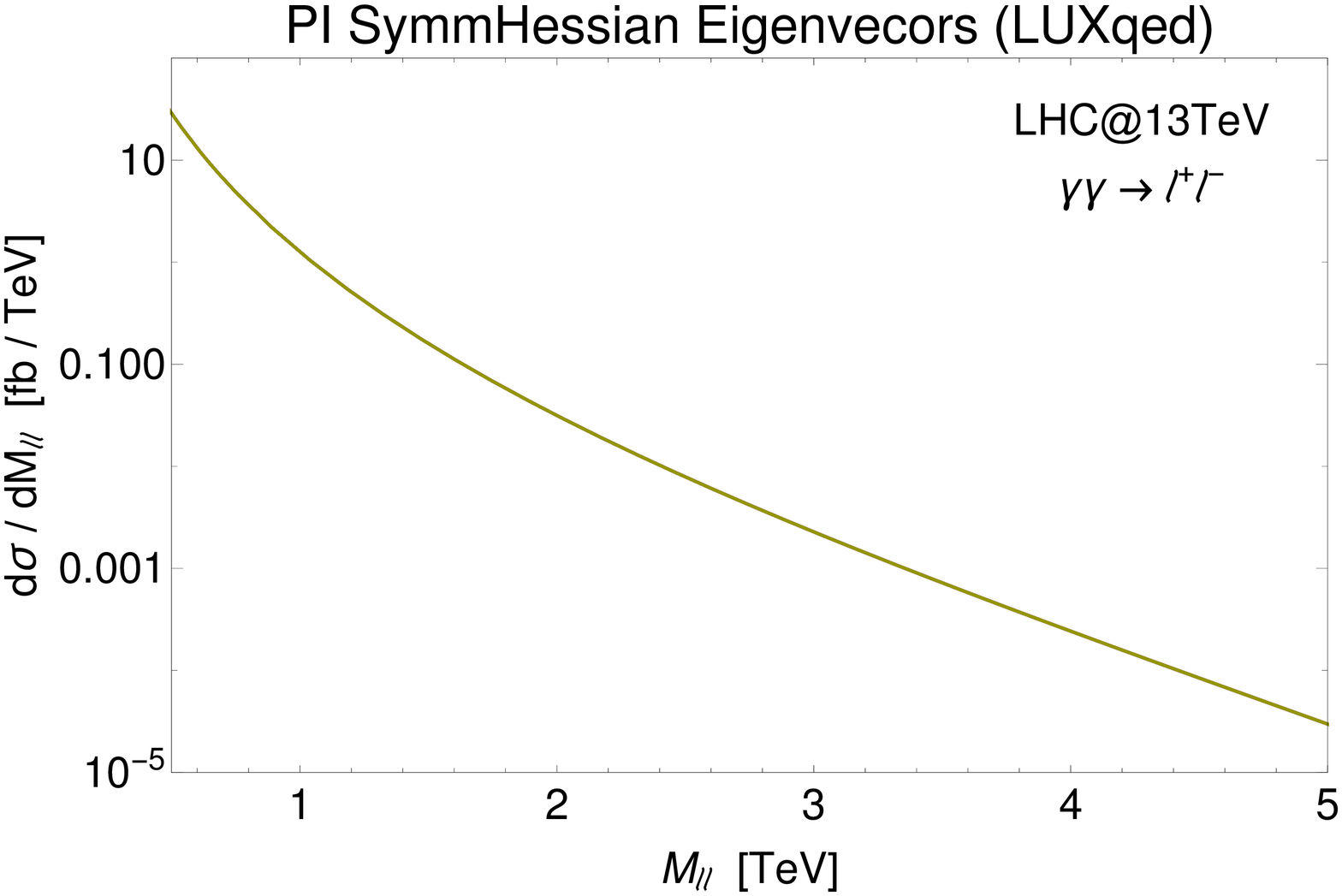}{\footnotesize (b)}
\caption{\footnotesize Inclusive PI predictions obtained with (a) the table of 31 PDFs of the CT14QED\_inc set and
(b) 100 Symmetric Hessian eigenvectors of the LUXqed set.}
\label{fig:replicas_hessian}
\end{figure}

\begin{figure}
\centering
\includegraphics[width=0.55\textwidth]{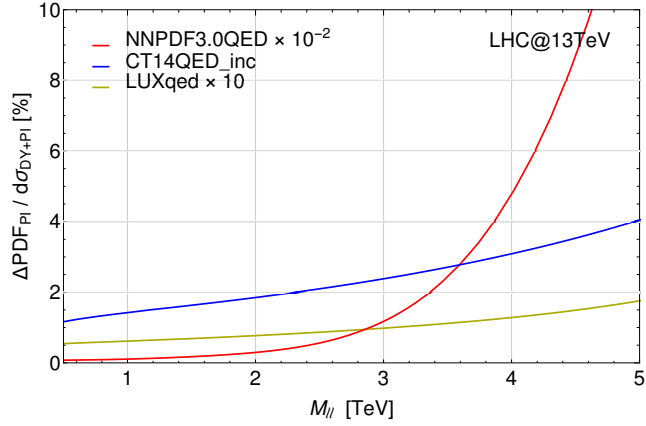}
\caption{\footnotesize Relative size of the photon PDF error over the complete di-lepton spectrum (PI+DY).
Note that the curves have been rescaled accordingly to the legend in order to fit in the plot.
}
\label{fig:PDF_error}
\end{figure}

In Fig.~\ref{fig:replicas_hessian} we show the predictions for the PI di-lepton spectrum obtained with the tables of the CT14QED\_inc set (a) 
and for the LUXqed Symmetric Hessian eigenvectors (b). 
From this picture we expect to obtain a small PDF uncertainty for the PI contribution.
This is shown in Fig.~\ref{fig:PDF_error}, where we plotted the impact of the photon PDF uncertainties on the total di-lepton spectrum i.e.,
the sum of the pure DY and the PI contributions.
The curves have been rescaled following the legend in order to be visible on the same scale.
The LUXqed predictions have the smallest uncertainty. 
The uncertainty on the di-lepton spectrum due to the photon PDFs is less than 0.2\%, thus well below any experimental uncertainty in 
the few TeV invariant mass region.
A slightly larger result is obtained from the CT14QED\_inc table, where the relative error from the photon PDF is between 1\% and 4\% in the 
same invariant mass interval.
The low uncertainty of the LUXqed and CT14QED\_inc predictions is shown in contrast with the high uncertainty predicted by the NNPDF3.0QED set.

\section{Photon-induced effects in $Z^\prime$-boson searches}

While narrow resonance searches in the di-lepton channel 
are not much affected by PI processes, in the case of broad 
$Z^\prime$ resonances, as well as in Contact Interaction (CI) scenarios, PI processes and 
their associated uncertainties may affect the sensitivity to new physics~\cite{Accomando:2016tah,Accomando:2016ehi}. 
The experimental approaches used to set limits on wide resonances (or CI) are essentially ``counting" strategies.
As these kinds of searches strongly rely on the good understanding of the background, the change of shape of the di-lepton spectrum
at high invariant masses due to the PI component has a crucial effect on the sensitivity.

\begin{figure}
\centering
\includegraphics[width=0.47\textwidth]{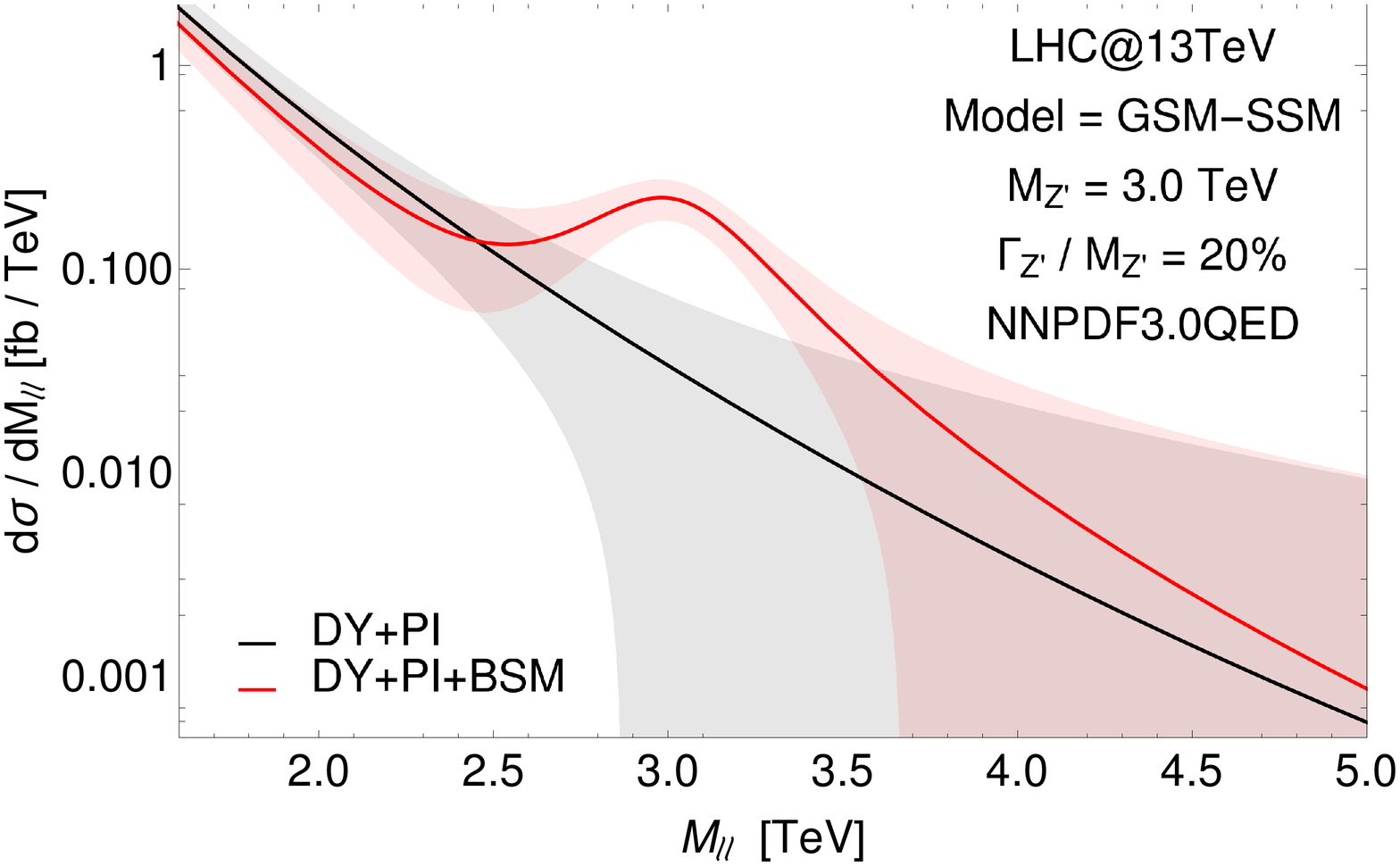}{\footnotesize (a)}
\includegraphics[width=0.47\textwidth]{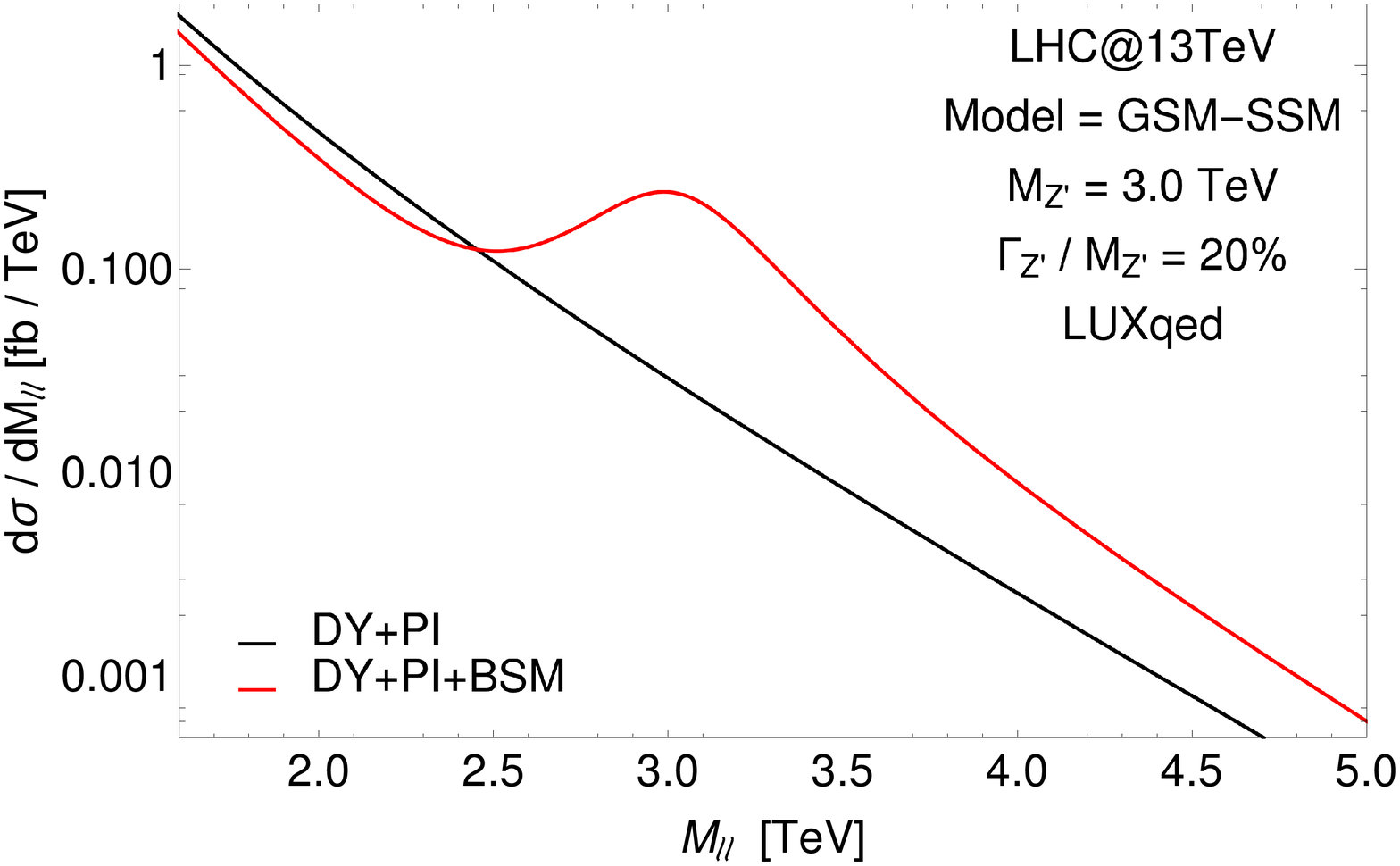}{\footnotesize (b)}
\includegraphics[width=0.47\textwidth]{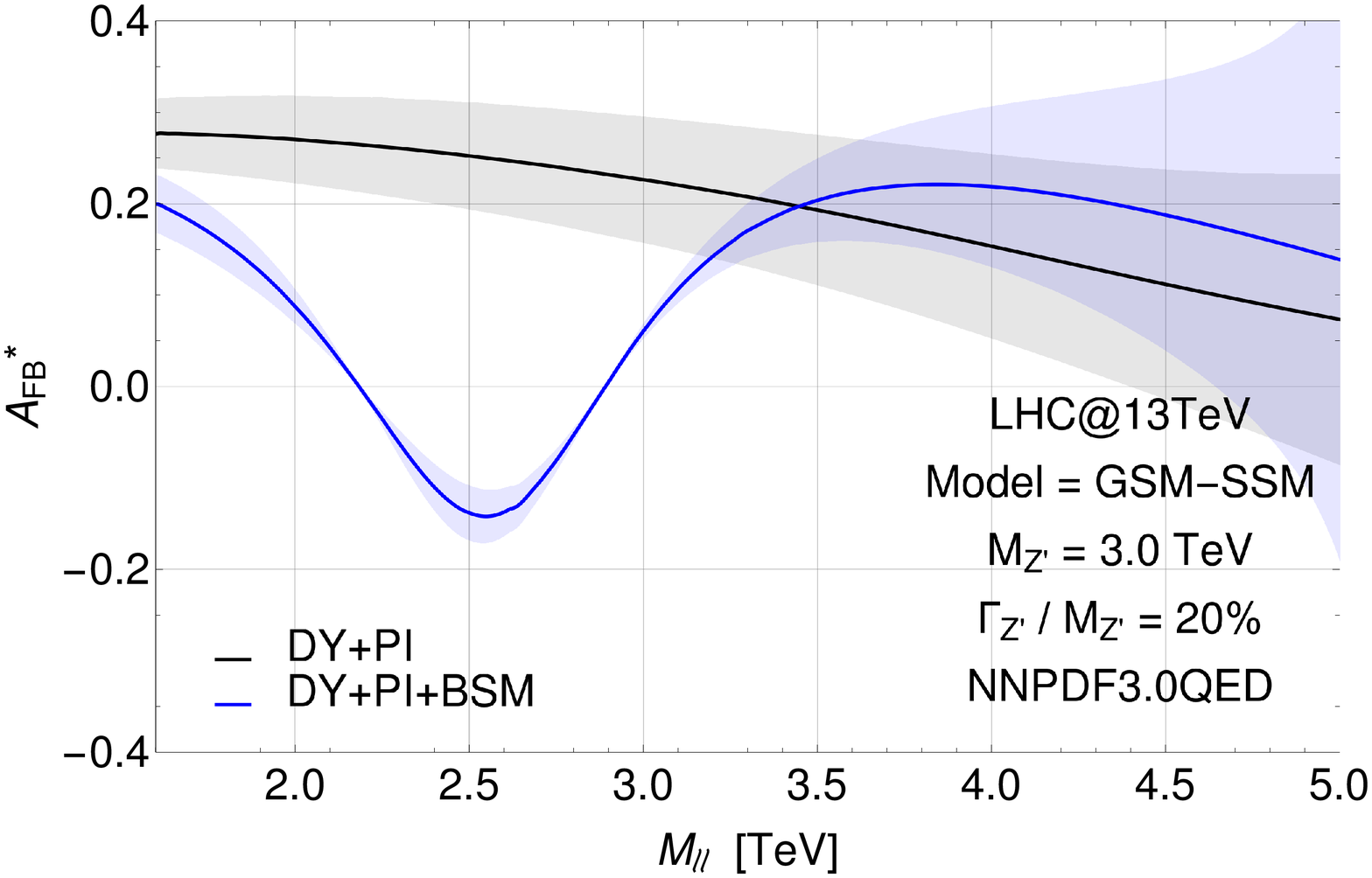}{\footnotesize (c)}
\includegraphics[width=0.47\textwidth]{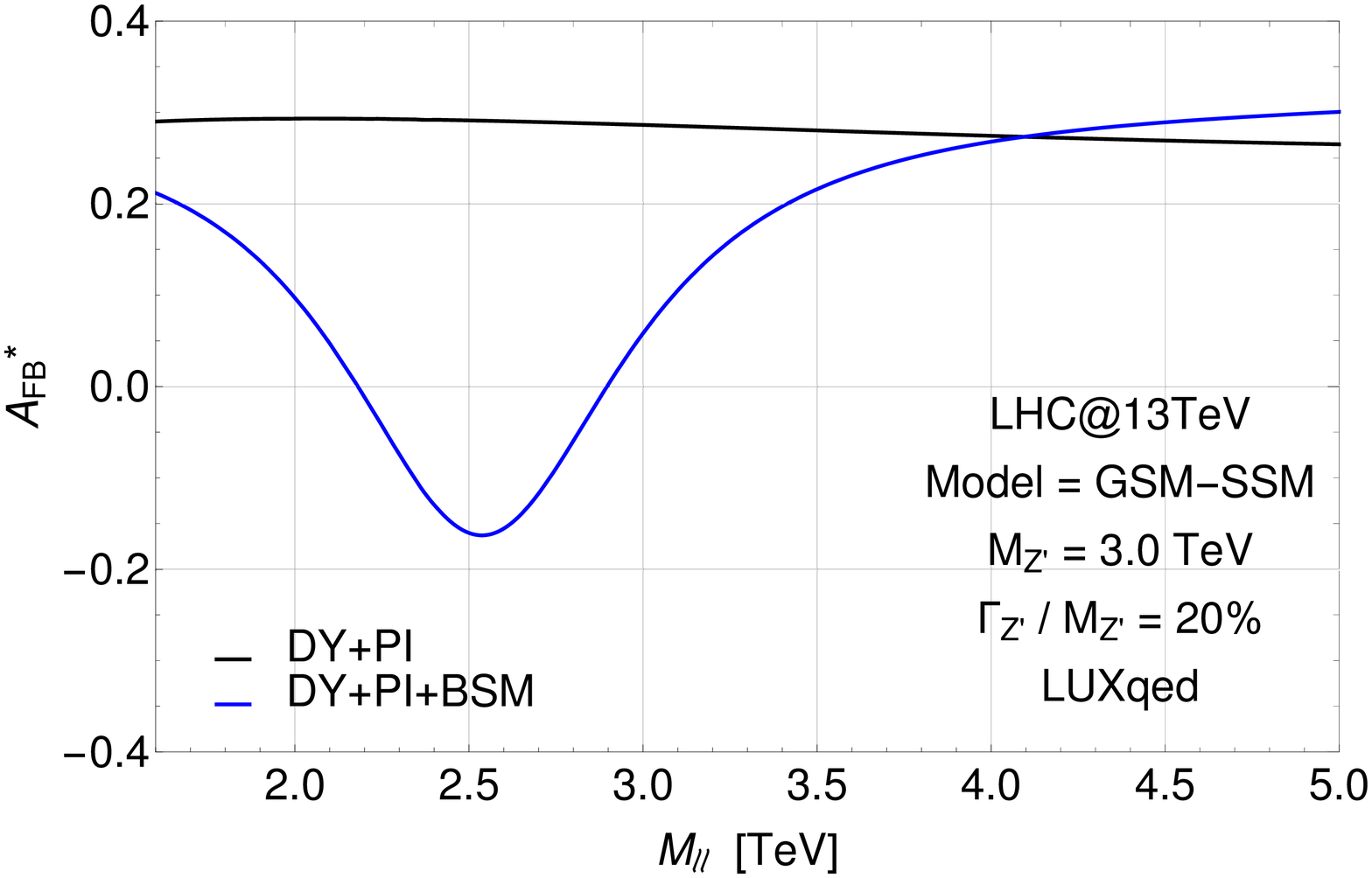}{\footnotesize (d)}
\caption{\footnotesize Differential cross section for the single $Z^\prime$ SSM benchmark with a mass of 3 TeV and $ \Gamma_{Z^\prime }/ M_{Z^\prime} = 20 \% $, with the PI contribution as predicted by the (a) NNPDF3.0QED and (b) LUXqed sets, including their photon PDF error. (c)-(d) Reconstructed AFB for the same benchmark.
}
\label{fig:Zprime}
\end{figure}

Fig.~\ref{fig:Zprime}~\cite{Accomando:2016ehi} shows the invariant mass 
spectrum of a single $Z^\prime$ Sequential Standard Model (SSM) benchmark as obtained from the 
NNPDF3.0QED and LUXqed PDF sets, and the reconstructed Forward Backward Asymmetry (AFB).
The latter has been considered because of its it features partial cancellation of PDF uncertainties, as shown in Ref.~\cite{Accomando:2015cfa}.
The NNPDF3.0QED scenario leads to a loss of sensitivity in the invariant mass spectrum, while the AFB maintains a visible shape standing over the systematics error bands 
even in the wide resonant scenario. The LUXqed low central value and small PDF uncertainty instead lead to no visible effects on both invariant mass 
and AFB distributions. 

\section{Conclusions}

The upgrade in energy of the LHC to 13 TeV at Run~II has opened the exploration of higher energy scales that were barred during the past Run~I. The LHC potential in BSM searches at the ongoing Run~II will be further boosted by the increase of the collected data sample in two years time when the luminosity should reach the project value $L=$ 300~fb$^{-1}$.
As we approach the high luminosity phase at the LHC, the statistical errors will be greatly reduced. At the same time, systematic effects will become more and more important.

One of the major sources of theoretical systematics at the LHC comes from PDF uncertainties. 
In this article we have studied the di-lepton final state in the 
high mass region $ M_{l l } \geq 1 $ TeV at the LHC Run~II 
and examined the impact of PDF systematics on searches for 
heavy neutral spin-1 $Z^\prime$-bosons, with particular regard to 
PI processes and associated effects of the photon PDF. 

Using as benchmarks the SM di-lepton spectrum and the GSM-SSM 
wide-resonance scenario, we have pointed out non-negligible contributions 
from the single-dissociative production process. This underlines the relevance of improving in the future the theory of 
PI processes at the LHC for both QED PDFs and elastic processes. 

We have analysed the significance of the BSM signal 
in di-lepton channels by incorporating real and virtual photon contributions. 
We have illustrated that this depends considerably on the different scenarios 
for photon PDFs, and presented results for the cross section and 
the reconstructed AFB as function of the di-lepton invariant mass. 

\vspace{-1em}
\section*{Acknowledgements}
\vspace{-1em}
E. A., J. F., S. M. \& C. S.-T. are supported in part through the NExT Institute.

\end{document}